# Revisiting THz Absorption in GO and rGO Liquid Crystalline Films

A. Vasil'ev[1, †], M. Zhezhu[1, †], H. Parsamyan[2, †], G. Baghdasaryan[1], M. Sargsyan[3, 4], D. A. Ghazaryan[3, 5], H. Gharagulyan[1,2]*

[1]*A.B. Nalbandyan Institute of Chemical Physics NAS RA, 5/2 P. Sevak str., Yerevan 0014, Armenia*
[2]*Institute of Physics, Yerevan State University, 1 A. Manoogian, Yerevan 0025, Armenia*
[3]*Laboratory of Advanced Functional Materials, Yerevan State University, Yerevan 0025, Armenia*
[4]*CANDLE Synchrotron Research Institute, 31 Acharyan St, Yerevan 0022, Armenia*
[5]*Moscow Center for Advanced Studies, Kulakova str. 20, Moscow, 123592, Russia*

*Author to whom correspondence should be addressed: herminegharagulyan@ysu.am
[†] These authors contributed equally to this work

**Abstract**

**With a swift progress in modern high-throughput communication systems, security, sensing and medicine utilizing THz range technologies, the demand for easy-to-fabricate, lightweight and high-performance absorbing materials has increased drastically. Notably, traditional approaches of eliminating unwanted radiation based on metasurfaces often face fabrication challenges limiting their practicality. In this study, we propose a straightforward approach for fabricating graphene oxide (GO) and reduced graphene oxide (rGO) liquid crystalline (LC) films *via* the vacuum filtration method and investigate their THz absorption characteristics. Here, the presence of LC phase in our electrochemically exfoliated GO and rGO LC films was confirmed by ellipsometric characterization. THz time-domain spectroscopy (TDS) measurements reveal that these films possess a low reflectance and transmittance confirming their strong absorptive properties within 0.4-1.6 THz frequency range for ~2 µm thick GO and rGO LC films. Particularly, the GOLC film shows ~37% average absorption at a thickness of 2.12 µm, which is 221 times smaller than the central wavelength. Similarly, the rGOLC film reaches ~50% absorption with a 1.68 µm thickness, 279 times smaller than the central wavelength. These findings provide valuable insights for development of GO- and rGO-based LC THz absorbers with highly tunable properties due to the ordering of GO flakes. Specifically, the LC phase of GO contributes to the formation of more uniform films with enhanced absorption due to the compact stacking and denser packing, compared to conventional GO films with randomly oriented GO flakes.**

**Abbreviations:** GO: graphene oxide; rGO: reduced graphene oxide; GOLC: graphene oxide liquid crystal; THz: terahertz; THz-TDS: terahertz time-domain spectroscopy.

# Introduction

The liquid crystalline (LC) phase of graphene oxide (GO) and reduced graphene oxide (rGO) has gained significant attention due to its self-assembly, tunable anisotropy, and potential applications in optoelectronics and biosensing recently [1, 2]. Flexible electronics based on the ordered thin films of these materials show great promise among these applications [3, 4]. In [5] the fabrication and characterization of graphene oxide liquid crystalline (GOLC) films with multiple layers is discussed highlighting its potential applications not only in flexible electrodes but also in electric circuits, chemical sensors, and functional anti-counterfeiting technologies. Free-standing GO and rGO platforms have also been successfully developed for various electrochemical and energy-related applications due to their excellent flexibility, mechanical stability, and conductivity [6-8]. These materials are typically fabricated via vacuum filtration of GO dispersions, often doped with functional molecules or composites, followed by chemical reduction to enhance performance [7]. Such rGO-based films demonstrate promising characteristics for use in wearable supercapacitors [1], electrochemical sensors [7], lithium-air cathodes [8] and batteries [9, 10] offering high capacitance, sensitivity, and structural integrity under repeated mechanical stress. Reduction of GO films into highly conductive and flexible graphene films that is emphasizing their potential for large-scale production is discussed in [11].

The typical methods for fabricating GO or rGO LC films from their dispersions include filtration, drop-casting, spin-coating, dip-coating, spray-coating, roll-to-roll coating, printing and solution spinning [12]. The technique of preparing GO-based films *via* vacuum filtration through a membrane offers several advantages. It allows precise control over the amount of GO solution used, does not require expensive or complex equipment, and eliminates the need for carrier gases to facilitate film deposition. Furthermore, the vacuum filtration method is highly scalable and does not involve any heating during the film deposition process. From a cost-wise standpoint, it is energy-efficient, which provides a clear advantage over the chemical bath deposition (CBD), chemical vapor deposition (CVD), shallow chemical bath deposition (SCBD) and spray pyrolysis methods [13-14]. Moreover, it is important to note that the method of electrochemical exfoliation of graphite results in a unique carbon product with functional groups that impart distinctive properties suitable for terahertz applications. In addition, CVD method yields graphene without functional groups [15].

These films demonstrate wide absorption and adjustable optical properties across various regions of the electromagnetic spectrum ranging from ultraviolet (UV) to terahertz (THz) ranges. Specifically, their optical characteristics are influenced by several factors, such as oxidation level, reduction state, and layer thickness [16]. As a result, GO and rGO LC films find extensive applications in optoelectronics, biomedical technologies, electromagnetic interference shielding, and communication systems [2, 17-20]. The challenges and perspectives of multifunctional films based on these materials is well discussed in [21]. In terahertz (THz) applications GO and rGO films are promising materials due to their unique structural and electronic properties [22, 23]. In particular, these films are ideal for applications in shielding, sensing and photonic devices [24-26]. Additionally, formation of GO and rGO LC phases strengthens their interaction with THz waves due to LC's self-assembling into structured anisotropic arrangements. From this point of view, the terahertz time-domain spectroscopy (THz-TDS) provides a powerful tool for fast and nondestructive characterization [27].

Designing a simple and highly efficient THz absorber with optimal absorptivity, bandwidth constraints, polarization insensitivity, and angular stability presents a significant challenge [28]. Specifically, current approaches have mainly focused on metamaterial absorbers, namely, designed metal-dielectric-metals structures [29], where a broadband absorption is traditionally achieved by combining multiple-sized subwavelength resonators or different structural layers in one unit-cell, thus, with antireflective properties and absorption resonances leading to narrowband operation, while some proposed structures incorporating multilayer graphene sheets which require complex design and fabrication techniques [30, 31]. Addressing these challenges through novel materials, innovative fabrication methods, and improved structural designs will be crucial for the future development of high-performance THz absorbers. THz absorbers based on GO and rGO films are extensively being investigated due to their unique physicochemical properties [32, 33]. In addition, the tunable and switchable THz absorption properties of GO and rGO LC films can be tailored by oxygen- containing functional groups, the degree of reduction, anisotropic alignment and modulation by external fields.

In this work, we thoroughly investigate GO and rGO LC films presenting thin and highly efficient broadband THz absorbers by using THz-TDS technique. In particular, electrochemically exfoliated GO and rGO LC films of different thicknesses were prepared and first characterized *via* XRD, Raman, and FTIR spectroscopy, as well as SEM and AFM microscopies. The presence of the LC phase in the structure was confirmed by ellipsometric characterization. It was shown that the metal-backed GO and rGO LC films with a thickness of approximately 2 µm can effectively absorb nearly 50 % of the incident THz radiation in the 0.4-1.6 THz spectrum due to their ordered structure and defective nature. A double layer of stacked rGOLC films will ensure absorption of 75 % with a central wavelength/thickness ratio of about 140. The latter is useful for a wide range of advanced technologies, including high-speed communication systems, stealth and biomedical applications. Thus, the LC phase of GO promotes the formation of uniform films with improved absorption, attributed to the compact stacking and higher packing density of GO flakes, in contrast to the randomly oriented structures typically observed in conventional GO films.

1. Materials and Methods

**Materials.** The graphite of > 99 % purity (CAS No.: 7782-42-5), gridded nitrocellulose filter (pore size: 0.45 µm), solvents, and all the chemicals used in this work were purchased from Sigma-Aldrich Chemical Co.

1.1. Characterization

Crystallographic data for GO and rGO LC films were obtained through XRD analysis (Rigaku MiniFlex instrument with CuKα – radiation) and its molecular conformations were analyzed *via* AFM-combined Raman Spectroscopy (LabRAM HR Evolution, HORIBA) using a 633 nm HeNe red laser with power of 0.5 mW. The chemical composition of the above-mentioned films was analyzed with FTIR-ATR Spectroscopy (Spectrum Two device, PerkinElmer). Birefringence in GO and rGO structures was investigated by Polarizing Optical Microscopy (MP920, BW Optics). Ellipsometric characterization of GO and GOLC films was done by Imaging Spectroscopic Ellipsometry (Accurion EP4, Park Systems).

1.2. THz Measurements

The THz spectral response of the material under test (MUT) was analyzed using an all-fiber-coupled THz time-domain spectrometer (TERA K15, Menlo Systems GmbH). The THz-TDS setup in both transmission and reflection configurations is presented in Fig. 1.

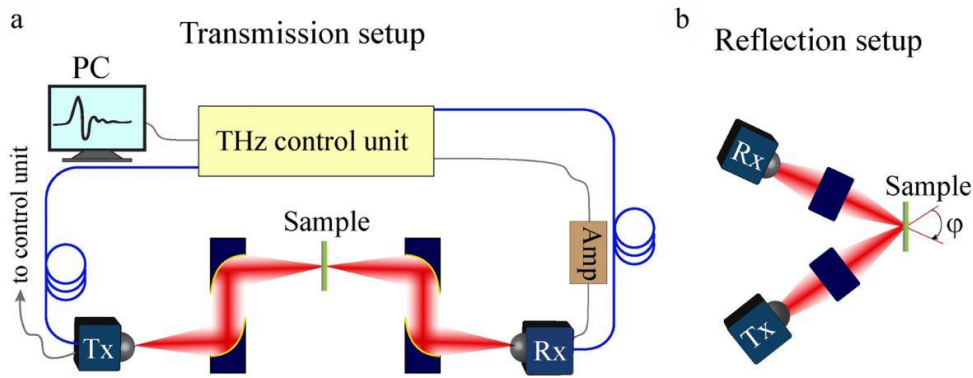

Fig. 1. Schematic representation of the experimental setup of the THz-TDS system for (a) transmission and (b) reflection measurements. $T_x$ and $R_x$ denote the transmitter and receiver antennas, respectively. The inset shows the sample holder with the material under the test (MUT). The measurements were carried out at room temperature.

Here, the THz signal was generated using a 1560 nm femtosecond laser pulse (100 MHz, 90 fs) with a dipole photoconductive antenna. For transmittance measurements, the linearly polarized THz wave was focused onto the sample using parabolic mirrors with a focal spot size of approximately 3 mm. The scanning range was set to 280 ps corresponding to a spectral resolution of about 1.83 GHz. The frequency-domain spectrum was obtained *via* Fourier transform of the detected transmitted signal with the transmission through air serving as the normalization reference. MUT was attached to a polypropylene foam polymer and positioned at the center of a rotatable sample holder between the antennas. Notably, polypropylene exhibits high transparency in the THz range. For the reflectance measurements a different measurement configuration was employed. The primary distinction from the transmission configuration was that the angle between the $T_x$ transmitter and $R_x$ receiver antennas was 70° corresponding to the incidence angle of 35° and the metallic mirror was placed behind the sample to facilitate metal-backed reflection recording.

### 1.3. GOLC and rGOLC Film Preparation

Electrochemical exfoliation of GO, its LC phase formation are well known and described in our previous works [34, 35]. Details of the electrochemical exfoliation procedure are provided in the Supplementary Materials. Approximately 30 mg of dry GO powder was obtained, corresponding to a yield of ~33%. Literature reports indicate that the yield of electrochemically exfoliated GO can range from 10% to 99%, depending on factors such as electrode type, electrolyte composition and concentration, and the applied electrochemical conditions [36]. Despite this variability, electrochemical exfoliation is widely favored due to its reduced processing time, environmental compatibility, and avoidance of the explosion hazards associated with traditional chemical oxidation methods [36-38].

The production of an LC structure in GO suspensions occurs through an entropy-driven process described by Onsager theory [39]. At low concentrations, GO sheets are randomly oriented (isotropic phase), but as the concentration or sheet aspect ratio increases, they begin to align parallel to minimize excluded volume and maximize packing entropy, forming a nematic phase [40]. This transition is also influenced by electrostatic repulsion generated from ionized carboxyl and hydroxyl groups on the GO sheet edges [41]. The formation of LC structures is additionally affected by factors such as pH, ionic strength, and the purity and size of the GO sheets [12, 35]. For the experiments, GOLC films were prepared by vacuum filtration method and reduced in hydrazine vapors (see Fig. 2). In particular, colloidal solution of GO with a concentration of 1.14 mg/ml was first prepared to obtain films of electrochemically exfoliated GO. The films were then deposited by filtering the GO solution through a nitrocellulose filter.

A film numbered as GOLC-1 was prepared from the original solution. The GO solution was then sequentially diluted to obtain films numbered GOLC-2, GOLC-3, GOLC-4, GOLC-5, and GOLC-6 yielding to varying thicknesses. The film thicknesses were assessed at 20 different regions *via* SEM and AFM techniques after which the results were averaged and the deviations were determined accordingly (see Table S1 in Supplementary Materials). The prepared films were used for further characterization and analysis.

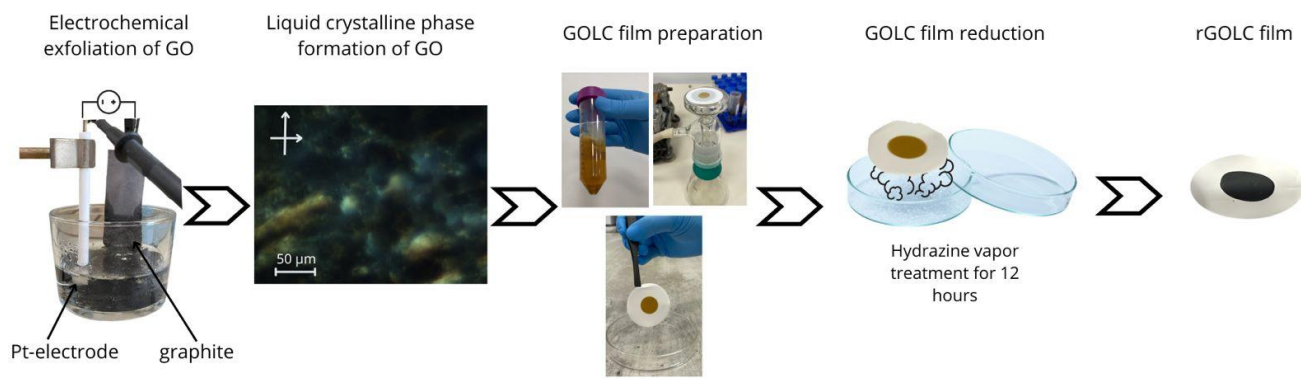

Fig. 2. Schematic representation of the GO synthesis and its LC phase formation: from the film preparation to the subsequent reduction procedure.

## 2. Results and Discussion

## 2.1. Ellipsometric Characterization of GO and GOLC films

To confirm the presence of LC phase, the birefringent properties of the GO and rGO structures were first studied by polarized optical microscopy (see in Supplementary Materials Fig. S1 (a) and (b)). Detailed ellipsometric characterization was then performed to compare the anisotropic properties of GO, rGO and GOLC films. For the determination of optical constants, measurements were carried out at room temperature. Monochromatic light was provided by a laser-driven Xenon lamp coupled to a grating monochromator. Measurements were performed over the accessible spectral range of 250–1000 nm with a 4 nm step, at the angles of incidence of 60°, 65°, and 70° with a 11x objective. Although the instrument offers high lateral resolution (< 1 µm), areas of at least 20 µm × 20 µm were measured to improve the accuracy. Here, multi-spot, multi-sample measurements were performed on films of varying thicknesses. A representative Δ and Ψ datasets for GO and GOLC films at a single location is shown in Supplementary Fig. S2, and for rGO films in Supplementary Fig. S3 (a).

Experimental data were fitted using various analytical models to extract each material's complex refractive index [42-44]. To minimize substrate-induced uncertainties, ellipsometric measurements were also taken on the bare glass substrate immediately adjacent to each film. The real part of the substrate's refractive index was modeled with the Cauchy dispersion relation: $n(\lambda)=A+B/\lambda^2+C/\lambda^4$, where λ is the probe wavelength, and A, B, and C are fitting parameters extracted from the Psi and Delta curves *via* the Levenberg–Marquardt algorithm. Surface roughness was incorporated into the model *via* the Bruggeman effective medium approximation [45].

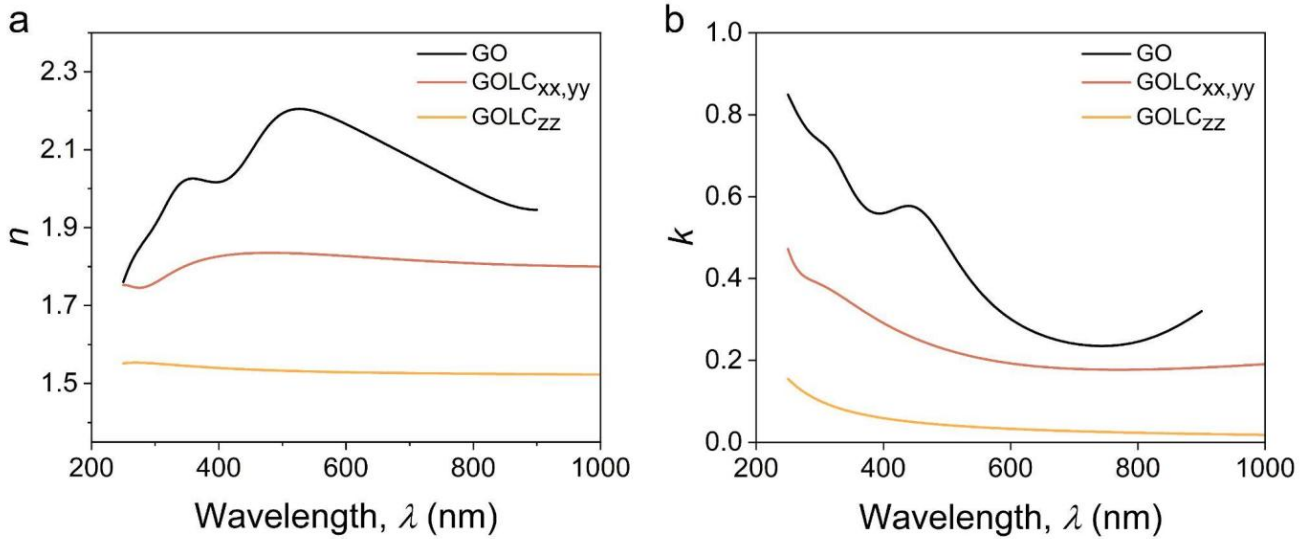

Fig. 3. Optical constants of GO and GOLC films obtained *via* spectroscopic ellipsometry. Real (a) and imaginary (b) parts of the refractive index as a function of wavelength, respectively.

The dielectric response of the GO films was described by an isotropic model [46] incorporating Lorentzian oscillators and given by $\varepsilon(E) = \varepsilon_\infty + \varepsilon_{UV}(E) + \sum_i \varepsilon_{L,i}(E)$. This model includes a high-energy limit $\varepsilon_\infty$ and a UV pole ($\varepsilon_{UV}$) term to capture absorption at higher photon energies. Surface roughness was also included in the analysis, and the mean squared error (MSE) was minimized as the goodness-of-fit criterion. The resulting Lorentzian-oscillator model fit is shown in Fig. S2, with an RMSE of 2.554. The extracted refractive index and extinction coefficient of GO are displayed in Fig. 3. Similar to an earlier report [44], an anisotropic model was applied to describe the optical properties of the studied GOLC films. The in-plane dielectric constant as a function of energy is expressed using Lorentzian and Gaussian oscillators, while a single oscillator is employed to capture the out-of-plane contribution to the dielectric constant:

$$\varepsilon_{IP}(E) = \varepsilon_\infty + \varepsilon_{UV}(E) + \varepsilon_L(E) + \sum_i \varepsilon_{G,i}(E), \tag{1}$$

$$\varepsilon_{OP}(E) = \varepsilon_\infty + \varepsilon_L(E). \tag{2}$$

The anisotropic model provides an overall good fit between experimental and modeled data (RMSE = 13.79), confirming the uniaxial optical properties of GOLC films. The extracted ordinary and extraordinary refractive indices and extinction coefficients of the GOLC films are depicted in Fig. 3. The results of ellipsometric characterization of rGO films are provided in Supplementary Materials (Fig. S3).

## 2.2. Structural Analysis of GOLC Films

Structural and spectral characterization of the GO and rGO LC films was conducted using a wide range of methods. In Fig. 4(a), images of the GOLC samples are shown arranged from the thickest GOLC-1 to the thinnest GOLC-6 and corresponding to the thicknesses of 2.12 µm, 1.71 µm, 1.22 µm, 0.72 µm, 0.49 µm, and 0.28 µm, as well as images of the rGOLC samples with thicknesses of 1.68 µm (rGOLC-1); 1.04 µm (rGOLC-2); 0.78 µm (rGOLC-3); 0.49 µm (rGOLC-4); 0.38 µm (rGOLC-5); 0.16 µm (rGOLC-6), respectively. The detailed surface morphology and topology analysis for the synthesized GO and rGO LC films were done *via* SEM and AFM techniques. As can be seen from SEM images, for the GOLC film the surface is more rippled and wrinkled due to its folded nature, as well as *sp³* centers and point defects (see Fig. 4 (b)), however, for the rGOLC film, a bubble-like structure forms leading to the increased surface roughness (see Fig. (c)). The same peculiarities were observed from the AFM analysis (see in the Supplementary Materials Fig. S4 (a) and (b), respectively). The root mean square (RMS) roughness of the GOLC-6 film, which exhibits the lowest thickness, constitutes approximately 33.4% of its total thickness. Following the reduction process, the RMS roughness of the resulting rGOLC film increases to approximately 84.9% of its thickness. The thicknesses of the GOLC samples numbered from GOLC-1 to GOLC-6 were defined (see Table S1) based on the cross-sectional SEM measurements. The same estimations were done for the rGOLC films with thicknesses ranging from 1.68 µm (rGOLC-1) to 0.16 µm (rGOLC-6). Here, Fig. (d) and (e) demonstrate cross-sectional SEM images of GOLC-1 and GOLC-6 films, while Fig. 4 (f) and (g) depict the images of rGOLC-1 and rGOLC-6 films.

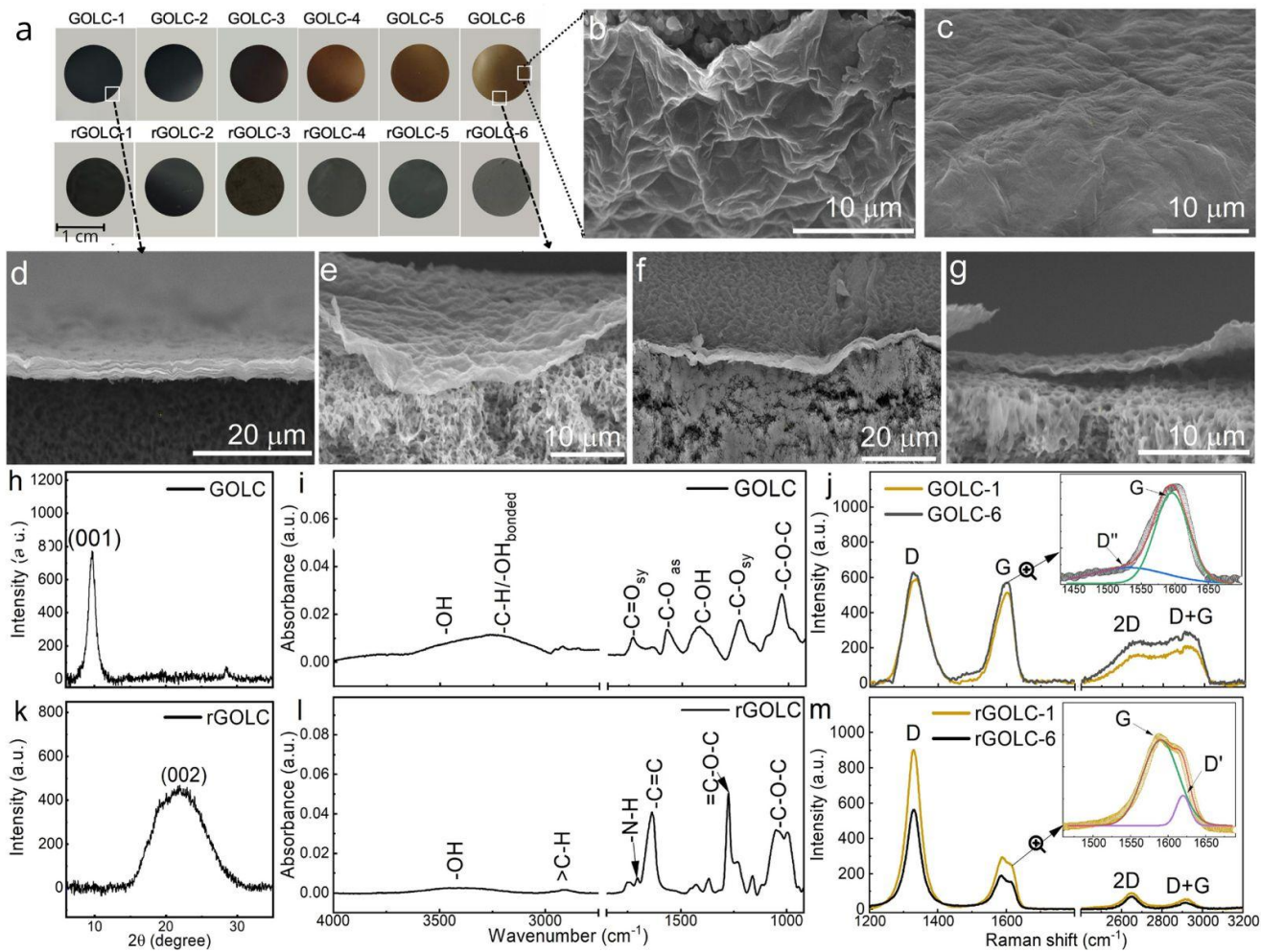

Fig. 4. Structural characterization of the GOLC and rGOLC films. (a) Pictures of GOLC samples with thicknesses: 2.12 μm (GOLC-1); 1.71μm (GOLC-2); 1.22μm (GOLC-3); 0.72μm (GOLC-4); 049 μm (GOLC-5); and 0.28 μm (GOLC-6), and rGOLC samples with thicknesses: 1.68 (rGOLC-1); 1.04 (rGOLC-2); 0.78 (rGOLC-3); 0.49 (rGOLC-4); 0.38 (rGOLC-5); 0.16 (rGOLC-6). SEM morphology images of (b) GOLC and (c) rGOLC films. Cross-sectional SEM images of GOLC films for (d) GOLC-1 and (e) GOLC-6. Corresponding cross-sectional SEM images of the rGOLC films for (f) rGOLC-1 and (g) rGOLC-6. Spectral characterization of the GOLC and rGOLC films. (h) XRD, (i) FTIR-ATR and (j) Raman spectra of the GOLC film for GOLC-1 and GOLC-6. (k) XRD, (l) FTIR-ATR and (m) Raman spectra of the rGOLC film reduced from the corresponding GOLC film for rGOLC-1 and rGOLC-6. All measurements were performed at a temperature of 25°C.

The spectral measurements for the GOLC and rGOLC films are depicted in Fig. 4. In particular, the XRD analysis was specifically conducted on GOLC-1 and rGOLC-1 to have stronger diffraction signals and better crystallinity detection. The diffraction peak of GOLC film appears at $2\theta = 9.66°$ (see Fig. 4 (h)), while for the rGOLC film, it is observed at $2\theta = 21.89°$ (see Fig. 4 (k)). The corresponding reflection planes are depicted in Supplementary Materials (see Table S2). The related interlayer spacings according to Bragg's equation ($d = \frac{\lambda}{2\sin\theta}$, where $\lambda$ is the wavelength of the X-ray source and $\theta$ is the angle of diffraction) are 0.91 nm and 0.41 nm, respectively. Consequently, the reduction process results in a 0.5 nm decrease in interlayer spacing. The interlayer spacing of GOLC is attributed to the presence of oxygenated functional groups and intercalated water molecules within the hydrophilic GO layers. The decomposition of the oxygen-containing groups on the graphene oxide sheets lead to the significant decrease in the interlayer spacing upon reduction of the GOLC film with hydrazine. The reduced spacing is close to that of natural graphite (0.334 nm) [47]. The crystallite size of GOLC was estimated from the XRD patterns using the Debye–Scherrer equation: $D = 0.9\lambda/(\beta \cos(\theta))$, where $\lambda$ is the X-ray wavelength 0.15418 nm, $\beta$ is the FWHM of the GO diffraction peak [48]. Based on this analysis, the average crystallite size was calculated to be approximately 6.5 nm. The number of graphene layers was then estimated by dividing the crystallite size by the interlayer spacing, yielding an average of about 7 layers. This value is consistent with the literature data for electrochemically synthesized graphene oxide [49].

The FTIR-ATR spectrum for GOLC-1 film is presented in Fig. 4 (i). The peaks at 1729 cm$^{-1}$, 1562 cm$^{-1}$, 1412 cm$^{-1}$, 1225 cm$^{-1}$, and 1163 cm$^{-1}$ correspond to vibrations of -C=O sy, -C-O as, C-OH, -C-O sy, and -C-O-C, respectively [50, 51]. The peaks corresponding to the vibrations of carbonyl, carboxyl, and hydroxyl groups showed very low intensity or disappeared entirely after the chemical reduction of GO films with hydrazine (see Fig. 4 (l)). The most intense peaks in the rGO spectrum at 1641 cm$^{-1}$, 1275 cm$^{-1}$, and 1050 cm$^{-1}$ correspond to -C=C, =C-O-C, and -C-O-C groups [47, 52]. The reduction of graphene oxide with hydrazine introduces N–H groups, as indicated by a minor peak at 1701 cm$^{-1}$ in the ATR spectrum of rGOLC (Fig. 4 (l)). In addition, the C-O-C groups are more dominant for the electrochemically exfoliated GO-deposited films. Carbonyl groups are primarily generated at the edges of GO layers, while the epoxy groups are located on the basal plane of GO [37]. A more detailed characterization of GO and rGO LC films using FTIR-ATR spectroscopy is provided in Supplementary Materials (see Table S2).

Fig. 4 (j) and (m) present Raman spectra of GOLC-1, GOLC-6 and rGOLC-1, rGOLC-6 films, respectively. The spectra exhibit four distinct peaks: *D*, *G*, *2D*, and *D + G*. The *D*-band is associated with edges, defects and structurally disordered carbons in graphene sheets [54], while the *G*-band corresponds to the vibration of *sp*$^2$ bonded carbon [55]. The *2D* band is a second-order overtone of in-plane vibrations, and the *D + G* band arises from the combined scattering of *D* and *G* modes [56]. The positions, line shapes, and intensities of these peaks provide valuable insights into the structural and electronic properties of these materials [57]. The Raman spectra of GOLC films exhibit a general trend where increasing film thickness results in a blue shift of *D*, *G*, and *2D* peaks. The most pronounced shift is observed in *2D* peak, with a maximum shift of Δ = 4.5 cm$^{-1}$ between GOLC-6 and GOLC-1. The Raman spectra of all GOLC and rGOLC films are presented in Fig. S5 with detailed peak position data provided in Table S3. Analysis of the spectra of GOLC films indicates that GOLC-5 and GOLC-6 exhibit a pronounced shoulder on the short-wavelength side of *G* peak. Gaussian fitting of this region revealed an additional peak at 1531 cm$^{-1}$, known in the literature as the *D″* peak. Some studies attribute the *D″* peak to amorphous carbon or interstitial defects linked to amorphous *sp*$^2$ bonded structures, which may incorporate functionalized small molecules [58, 59].

After the chemical reduction process, the Raman spectra of the rGOLC film exhibit significant changes in line shape. This is logical as chemical reduction induces a significantly higher number of defects compared to thermal reduction, which in its turn influences the physical and chemical properties of these materials. Furthermore, the reduction process with hydrazine as a reducing agent often leads to incorporation of functional groups onto GO sheets including amino nitrogen atoms [59, 60]. A comparison of the Raman spectra of rGOLC films reveals a narrowing of *D*-band and a decrease in *G*-band intensity accompanied by broadening. The narrowing of *D* peak is associated with a reduction in *sp*$^3$-like defects caused by deoxygenation of functional groups during the reduction process. However, the persistent high intensity of the *D*-band in rGOLC films suggests the presence of *sp*$^3$ carbon atoms and defect sites [47]. The decrease in *G* peak intensity is accompanied by the band splitting. For more precise analysis, the *G* peak region (1450-1700 cm$^{-1}$) was additionally fitted using a Gaussian function. The appearance of a peak at 1620 cm$^{-1}$ corresponds to the *D′* band, which has been attributed to double vacancies forming pentagonal and octagonal rings commonly referred to as 5-8-5 defects [54, 59], a characteristic feature of few-layer wrinkled systems [34, 61]. Additionally, the *D*, *2D*, and *D + G* peaks exhibit a shift toward the shorter-wavelength region. Notably, the most significant shift between GOLC-1 and rGOLC-1 (Δ = 68 cm$^{-1}$) is observed for *2D* peak, whose decreasing position can predict an increasing electron concentration in the system [62]. At the same time, the increased prominence of the *2D* and *D + G* peaks indicates reduced disorder, consistent with the removal of oxygen-containing functional groups. [57, 59]. This suggests that oxygen-containing functional groups and their associated defects were effectively eliminated following the reduction process.

However, the graphene lattice retains a significant number of defects, likely associated with vacancy formation, new crystalline boundaries, and the presence of non-standard rings with varying numbers of carbon atoms, leading to the generation of more point-like and line-like defects within the graphitic lattice [59, 63, 64]. The intensity ratio ($I_D/I_G$) was further used to assess the density of the small graphene domains and is presented for all films in Table 1. The chemical reduction process leads to a significant increase in this ratio: the $I_D/I_G$ ratio of rGOLC films rises approximately 2.5 times compared to GOLC films indicating that more $sp^2$ domains are formed and the defects on rGOLC are increasing [47].

To estimate defect distance ($L_D$) and density ($n_D$), we used equations (3) and (4) as in [65].

$$L_D^2(nm^2) = (1.8 \pm 0.5) \times 10^{-9} \lambda^4 \left(\frac{I_D}{I_G}\right)^{-1} \tag{3}$$

$$n_D(cm^{-1}) = \frac{(1.8 \pm 0.5) \times 10^{22}}{\lambda^4} \left(\frac{I_D}{I_G}\right) \tag{4}$$

where $\lambda$ is the laser source wavelength, and the $I_D/I_G$ is the intensity ratio of the *D* to *G* bands.

Table 1. The $I_D/I_G$ ratio, $L_D$ and $n_D$ for GOLC and rGOLC films.

| Film Number | GOLC | | | rGOLC | | |
|---|---|---|---|---|---|---|
| | $\frac{I_D}{I_G}$ | $L_D$, nm | $n_D$, $10^{11}$ cm$^{-1}$ | $\frac{I_D}{I_G}$ | $L_D$, nm | $n_D$, $10^{11}$ cm$^{-1}$ |
| 1 | 1.14 | 15.9±8.4 | 1.3±0.3 | 2.70 | 10.3±5.5 | 3.0±0.8 |
| 2 | 1.21 | 15.5±8.2 | 1.4±0.3 | 2.61 | 10.5±5.5 | 2.9±0.7 |
| 3 | 1.26 | 15.2±8.0 | 1.4±0.3 | 2.73 | 10.3±5.4 | 3.1±0.8 |
| 4 | 1.14 | 15.9±8.4 | 1.3±0.3 | 2.48 | 10.8±5.7 | 2.8±0.7 |
| 5 | 1.06 | 16.5±8.7 | 1.2±0.3 | 2.67 | 10.4±5.5 | 3.0±0.7 |
| 6 | 1.08 | 16.3±8.6 | 1.2±0.3 | 2.63 | 10.5±5.5 | 3.0±0.7 |

The defect distance for GOLC films is higher in GOLC-5, and it is the lowest for GOLC-6. After chemical reduction, the defect distance decreases approximately 1.5 times and the defect density nearly doubles. These results indicate that hydrazine reduction of GOLC films leads to an increase in the number of defective domains and a decrease in their size. This can likely be explained by the formation of unsaturated and conjugated carbon atoms during hydrazine treatment leading to the creation of new graphitic domains that are smaller than those present in GO before reduction, but numerous in quantity [66, 67]. These findings were further supported by the estimated $L_D$ and $n_D$ values. For GOLC-1 film $L_D = 15.9$ and $n_D = 1.3 \times 10^{11}$, while for rGOLC-1 one it is: $L_D = 10.3$ and $n_D = 3 \times 10^{11}$. For GOLC-6 film $L_D = 16.3$ and $n_D = 1.2 \times 10^{11}$ and for rGOLC-6: $L_D = 10.5$ and $n_D = 3 \times 10^{11}$.

## 2.3. THz-TDS Analysis of GOLC and rGOLC Films

The spectral characteristics of THz radiation absorption and its efficiency in GO and rGO LC films is governed by a complex interplay of their electronic, structural, and morphological factors. It is highly influenced by the degree of reduction, structural ordering, film thickness and the alignment of GO sheets with respect to each other. In graphitic materials, THz absorption primarily arises from free-carrier absorption, π–π electronic transitions or plasmonic resonances [25, 68].

THz spectral characterization of MUT consists of two steps: (1) a reference measurement without the sample, capturing the spectral response of the nitrocellulose filter used as a reference for normalization, and (2) the measurement of the sample as illustrated in Fig. 1. The absorbance is estimated based on transmittance (*T*) and reflectance (*R*) through the following relation:

$$A = 1 - T - R \tag{5}$$

However, since we measured the metal-backed reflectance (see Fig. 1), the incident THz wave first passes through the film, reflects off the metal surface, and then interacts with the film again, resulting in doubling of the absorption. Therefore, the absorbance was calculated using the following equation:

$$A = \frac{I_{abs}}{I_{ref}} = \frac{(I_{ref} - I_{samp})}{2 I_{ref}}, \tag{6}$$

where $I_{abs}$ is the absorbed intensity, $I_{ref}$ and $I_{samp}$ denote the intensities of the reference and the sample, respectively.

Fig. 5 (a) shows THz absorbance spectra for GOLC films with different thicknesses (from the thickest: GOLC-1 with 2.12 μm thickness to the thinnest: GOLC-6 with 0.28 μm thickness). It can be seen from the spectra that absorbance generally increases with frequency for all samples within the 0.4 - 1.6 THz range except for GOLC-6, which shows a slight deviation between 0.6 and 0.8 THz due to the film's inhomogeneity. Here, the absorbance varies from 8.6 % to 17.1 %. The thicker films exhibit higher absorbance as expected. In particular, the absorbance of GOLC-1 increases from 26.9 % to 41.1 % in the 0.4 - 1.6 THz frequency range. In Fig. 5 (b) similar spectra are presented for rGOLC films. After hydrazine reduction, the absorbance of rGOLC-6 remains nearly constant compared to GOLC films within the range of 8.8 % to 28.3 %, whereas for rGOLC-1, a significant increase is observed with absorbance values rising to 48.0 % - 49.9 %. rGOLC-1 exhibits almost constant absorbance across the entire spectrum of interest possibly due to saturation of the electronic transitions or a dense packing. rGOLC films show higher absorbance than the corresponding GOLC films, potentially due to increased defect density resulting from the reduction. Namely, the increased absorption in rGOLC films can be attributed to the fact that although most of the functional groups diminish after reduction, lattice distortion occurs on the basal plane leading to the formation of new defects. These defects create additional absorption sites, thereby enhancing the overall absorption properties of these films as previously observed in Raman studies. On the other hand, THz radiation absorption of the films is highly influenced by the arrangement of *sp²*-hybridized carbon, carbon content, and the size of graphitic domains [69]. The restoration of *sp²*-hybridized domains during the reduction enhances the delocalization of π-electrons leading to the increased conductivity and stronger interaction with THz waves.

Fig. 5 (c) shows the temporal waveform of the amplitude of a THz pulse for GOLC-1 and rGOLC-1 films and a nitrocellulose filter as a reference. The comparison of temporal profiles of THz pulses clearly shows the absorption behavior of the films. The THz temporal responses of the thinner films are shown in Supplementary materials (see Fig. S6). The histogram in Fig. 5 (d) illustrates the absorbance ratio of the GOLC and rGOLC films as a function of film thickness. Here, the highest values of the absorption ratio were observed for sample 4 with the thicknesses of 0.72 μm (GOLC) and 0.49 μm (rGOLC).

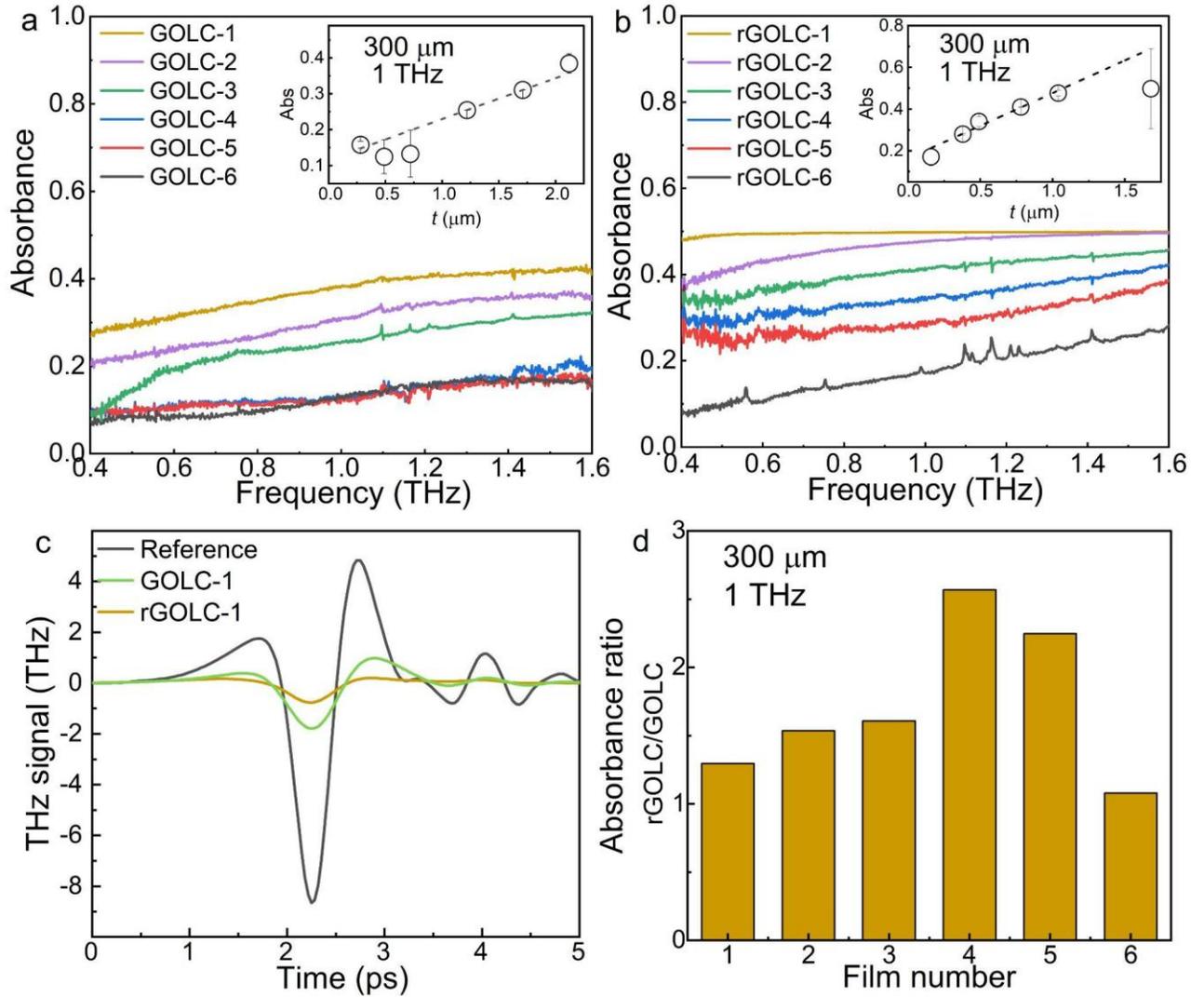

Fig. 5. THz response of GOLC and rGOLC films. Absorption spectra of (a) GOLC and (b) rGOLC films with different thicknesses. Insets show absorption coefficient dependencies on the film thickness. (c) Temporal profiles of a THz pulse for GOLC-1 and rGOLC-1 films and the reference filter. (d) Histogram of the absorbance ratio (rGOLC/GOLC) as a function of film thickness for GOLC-1 and rGOLC-1; GOLC-2 and rGOLC-2; GOLC-3 and rGOLC-3; GOLC 4 and rGOLC-4; GOLC-5 and rGOLC-5; GOLC-6 and rGOLC-6, respectively.

Transmission spectra of the same films were also recorded to confirm that the observed signal is due to actual absorbance rather than reflection or transmission effects (see Fig. S7 in Supplementary materials). Transmission coefficients were obtained by the following equation:

$$T(dB) = 10 \times \left(\frac{P_t}{P_i}\right) = SE\ (dB), \qquad (7)$$

where $P_t$ is transmitted and $P_i$ the incident powers; $SE$ is shielding efficiency [54]. The transmittance of GOLC-6 ranges from 88.8 % to 83.4 % across the 0.4-1.6 THz frequency range, whereas for the GOLC-1, it decreases from 61.8 % to 27.2 %. After chemical reduction, the transmittance of rGOLC-6 drops to a range of 54.2 % to 46.4 %, while that of rGOLC-1 falls further, remaining between 4.5 % and 1.7 % within the same frequency range.

Table 2 presents a comparison of THz absorption characteristics of our synthesized GOLC films with the graphitic films reported in the literature. For comparison, the absorption range, absorbance percentage (thickness) and $\lambda_0$ to thickness ratio were chosen as criteria, where $\lambda_0$ is the central wavelength of the reported absorption band. The rGO paper, rGO film and 3D graphene were selected for comparison. It is seen that nearly 37 % average absorption is achieved by the GOLC film with a thickness of about 2.12 μm, which is 221 times smaller than the center wavelength of the absorption band. Similarly, the average absorption of the rGOLC film is approximately 50 % with a thickness of 1.68 μm - nearly 279 times smaller than the center wavelength of the absorption band. Considering these values, a double layer of stacked rGOLC films will ensure absorption of 75 % with a $\lambda_0$/thickness ratio of about 140. Additionally, a broadband absorption is obtained using a straightforward and easily implementable film fabrication method.

Table 2. Comparison of the performance of the proposed absorber with other THz absorbers.

| Materials | Frequency range | Absorbance | Thickness | $\lambda_0$/thickness | Ref |
|---|---|---|---|---|---|
| rGO paper | 0.2-1.2 THz | ~ 64 % | 13 μm | 67 | [24] |
| rGO paper | 0.1-0.8 THz | 98% at 0.7 THz | ~370 μm | 4.6 | [33] |
| rGO film | 2.52 THz | 75 - 95 % | 1.2-1.5 μm | 88 | [68] |
| 3D graphene | 0.15-10 THz | 85 % | 1.6 mm | 0.5 | [71] |
| GOLC film | 0.4-1.6 THz | 27 – 41 % | 2.12 μm | 221 | This work |
| rGOLC film | 0.4-1.6 THz | 49 - 50 % | 1.68 μm | 279 | This work |

The prepared films demonstrated stable THz absorption characteristics over 6 months when stored under ambient conditions, indicating good stability. Besides, multiple samples were fabricated under identical conditions, and the THz absorption responses showed consistent performance with minimal variation, confirming the reproducibility of the method.

**Conclusion**

In this study, we demonstrated the potential of GOLC and rGOLC films as efficient THz absorbers. Particularly, GOLC and rGOLC films with varying thicknesses were prepared from electrochemically exfoliated water dispersions, then analyzed using XRD, Raman, and FTIR-ATR spectroscopy, along with SEM and AFM microscopy. Thorough ellipsometric characterization was done for confirmation of the existence of LC phase in GO structure. THz-TDS measurements confirm strong absorption in the 0.4-1.6 THz range by GOLC and rGOLC films, namely the GOLC film achieves an average absorption of ~37% at a thickness of 2.12 μm, which is 221 times smaller than the central wavelength of the absorption band. The rGOLC film shows ~50% average absorption with a 1.68 μm thickness, corresponding to a central wavelength/thickness ratio of 279. Based on these results, stacking two rGOLC layers could yield ~75% absorption with a central wavelength/thickness ratio of approximately 140. The LC phase of GO facilitates the formation of more uniform films with improved absorption, owing to tighter stacking and denser packing of the flakes, in contrast to conventional GO films composed of randomly oriented sheets. Raman analysis reveals that the enhanced absorption in rGOLC films arises from defect formation due to lattice distortion after reduction despite the decrease in functional groups. These findings highlight the effectiveness of GO and rGO-based LC films for tunable THz absorption paving the way for their application in advanced THz technologies.

# Supplementary Materials for
# Revisiting THz Absorption in GO and rGO Liquid Crystalline Films

A. Vasil'ev[1, †], M. Zhezhu[1, †], H. Parsamyan[2, †], G. Baghdasaryan[1], M. Sargsyan[3, 4], D. A. Ghazaryan[3, 5], H. Gharagulyan[1,2,]*

[1]A.B. Nalbandyan Institute of Chemical Physics NAS RA, 5/2 P. Sevak str., Yerevan 0014, Armenia
[2]Institute of Physics, Yerevan State University, 1 A. Manoogian, Yerevan 0025, Armenia
[3]Laboratory of Advanced Functional Materials, Yerevan State University, Yerevan 0025, Armenia
[4]CANDLE Synchrotron Research Institute, 31 Acharyan St, Yerevan 0022, Armenia
[5]Moscow Center for Advanced Studies, Kulakova str. 20, Moscow, 123592, Russia

*Author to whom correspondence should be addressed: herminegharagulyan@ysu.am
† These authors contributed equally to this work

**GO Synthesis**

Electrochemical exfoliation was performed in two steps using a standard cell with a platinum cathode and graphitic paper anode (2 × 4 cm) placed 2 cm apart. In the first step, 98% sulfuric acid (∼100 mL) was used as the electrolyte to enable intercalation of $HSO_4^-$ anions into the graphite structure. The process was conducted in galvanostatic mode with a specific current density of 50 mA/cm$^3$ until the cell voltage reached 2.2 V. The blue coloration of the graphite surface indicated that intercalation had occurred successfully and signaled the transition to the second step.

For the second stage, the electrodes were transferred to a 0.1 M ammonium sulfate (($NH_4)_2SO_4$) electrolyte (∼75 mL), where exfoliation proceeded under potentiostatic conditions at an applied voltage of 15 V. The process proceeded until the graphitic paper broke apart, resulting in the formation of a colloidal suspension of electrochemically exfoliated GO.

The resulting GO suspension was filtered, diluted with approximately 50 mL of distilled water, sonicated for 30 minutes, and centrifuged. The final GO product was dried using freeze-drying method to obtain dry GO powder.

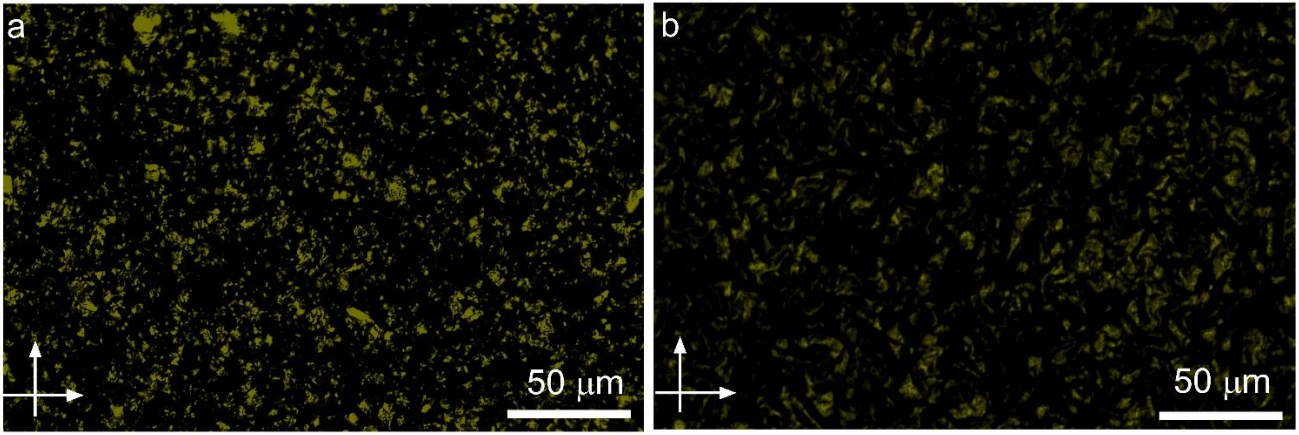

Fig. S1. Typical optical micrographs of liquid crystalline phases of (a) GO and (b) rGO films between crossed polarizers.

Supplementary Fig. S2 displays the experimental and fitted ellipsometric parameters for the GO and GOLC films studied in the main text.

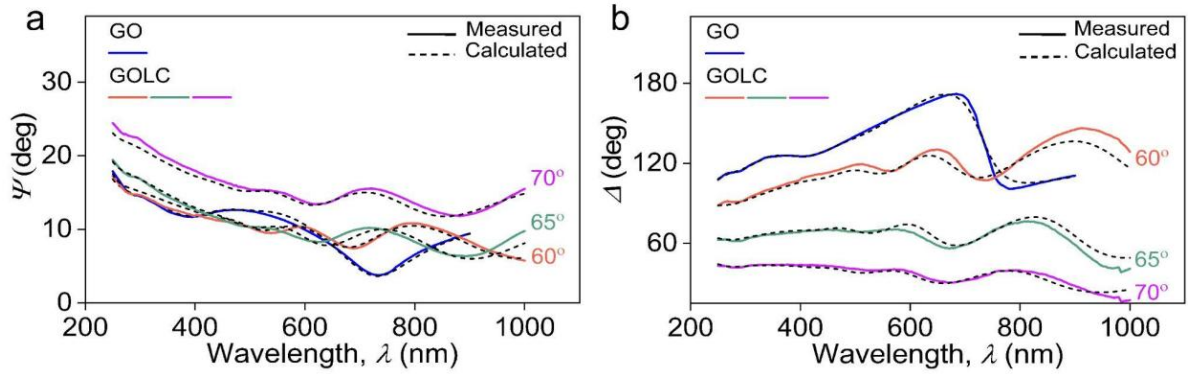

Fig. S2. The experimental and fitted ellipsometric parameters of GO and GOLC films at different incidence angles. Δ and Ψ values for GO films (blue lines) were measured at a 60° angle of incidence and fitted with the Lorentzian-oscillator model. Δ and Ψ values for GOLC films were measured at three different incidence angles and subsequently fitted with the anisotropic model described in the main text.

**Ellipsometric Characterization of Reduced Graphene Oxide (rGO) Films**

Measurements were performed over the accessible spectral range of 250–1000 nm, at angles of incidence of 60° and 70°. An isotropic model, incorporating Lorentzian oscillators and given by

$$\varepsilon(E) = \varepsilon_\infty + \sum_i \varepsilon_{L,i}(E),$$

was applied to describe the optical properties of the studied rGO films. This model also includes $\varepsilon_\infty$ to define the dispersion's high-energy limit. The ellipsometry data and corresponding model fits (RMSE = 2.684) are shown in Fig. S3a. The extracted optical constants are shown in Fig. S3b.

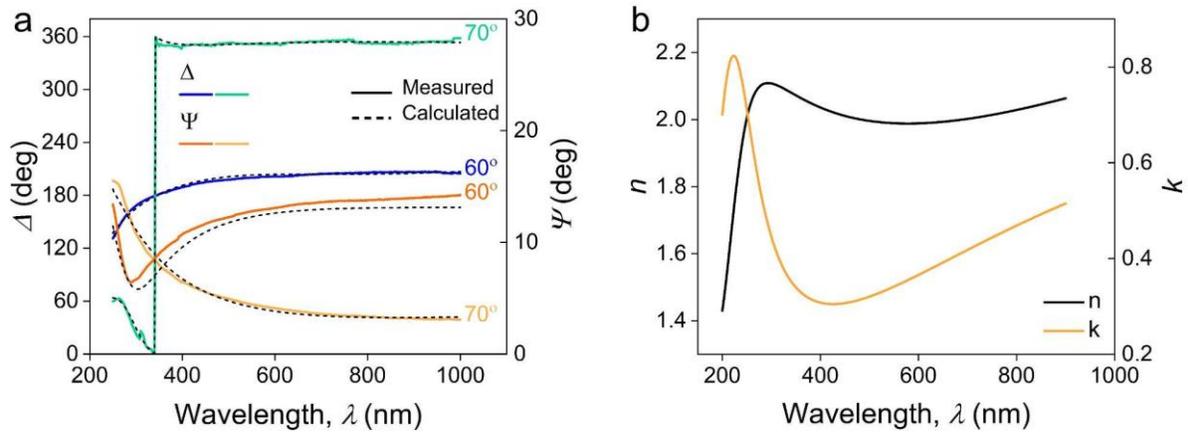

Fig. S3. Ellipsometric characterization of the rGO films. (a) Measured and fitted ellipsometric parameters at two incidence angles. (b) Extracted optical constants for rGO films.

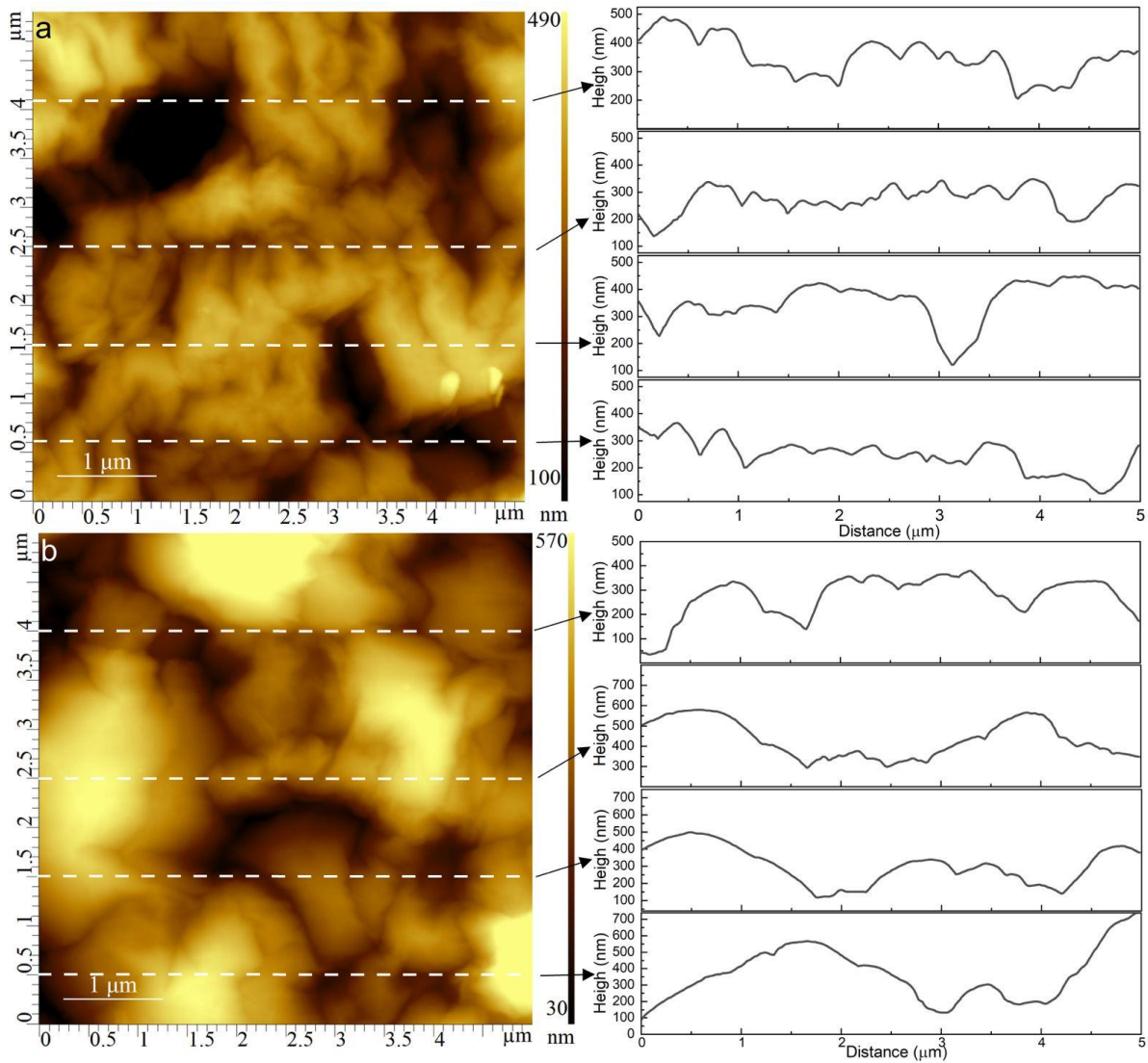

Fig. S4. AFM images of (a) GOLC and (b) rGOLC films and their roughnesses.

Table S1. Thicknesses of GOLC and rGOLC films.

|   | GOLC, $\mu$m |   | rGOLC, $\mu$m |
|---|---|---|---|
| 1 | 2.12 ± 0.17 | 1 | 1.68 ± 0.26 |
| 2 | 1.71 ± 0.35 | 2 | 1.04 ± 0.34 |
| 3 | 1.22 ± 0.35 | 3 | 0.78 ± 0.28 |
| 4 | 0.72 ± 0.10 | 4 | 0.49 ± 0.09 |
| 5 | 0.49 ± 0.18 | 5 | 0.38 ± 0.07 |
| 6 | 0.28 ± 0.05 | 6 | 0.16 ± 0.04 |

Table S2. Characterization of GOLC-1 and rGOLC-1 films by XRD analysis and FTIR-ATR Spectroscopy.

| XRD | | |
|---|---|---|
| | 2$\theta$, degree | Reflection plane |
| GOLC-1 | 9.66 | (001) |
| rGOLC-1 | 21.89 | (002) |
| FTIR | | |
| GOLC-1 | rGOLC-1 | |
| Wavenumber, cm$^{-1}$ | Wavenumber, cm$^{-1}$ | Bands |
| 3286 | 3400 | -OH, -C-H/-OH bonded |
|  | 2912 | >C-H stretch |
| 1729 |  | -C=O sy |
|  | 1701 | -N-H |
| 1645 | 1641 | -C=C |
| 1562 | - | -C-O as |
| 1412 | 1430 | C-OH |
|  | 1369 |  |
| - | 1275 | =C-O-C |
| 1225 | 1224 | -C-O sy |
| 1163 | 1163 | -C-O-C basal sy |
| 1034 | 1050 | -C-O-C edge sy |

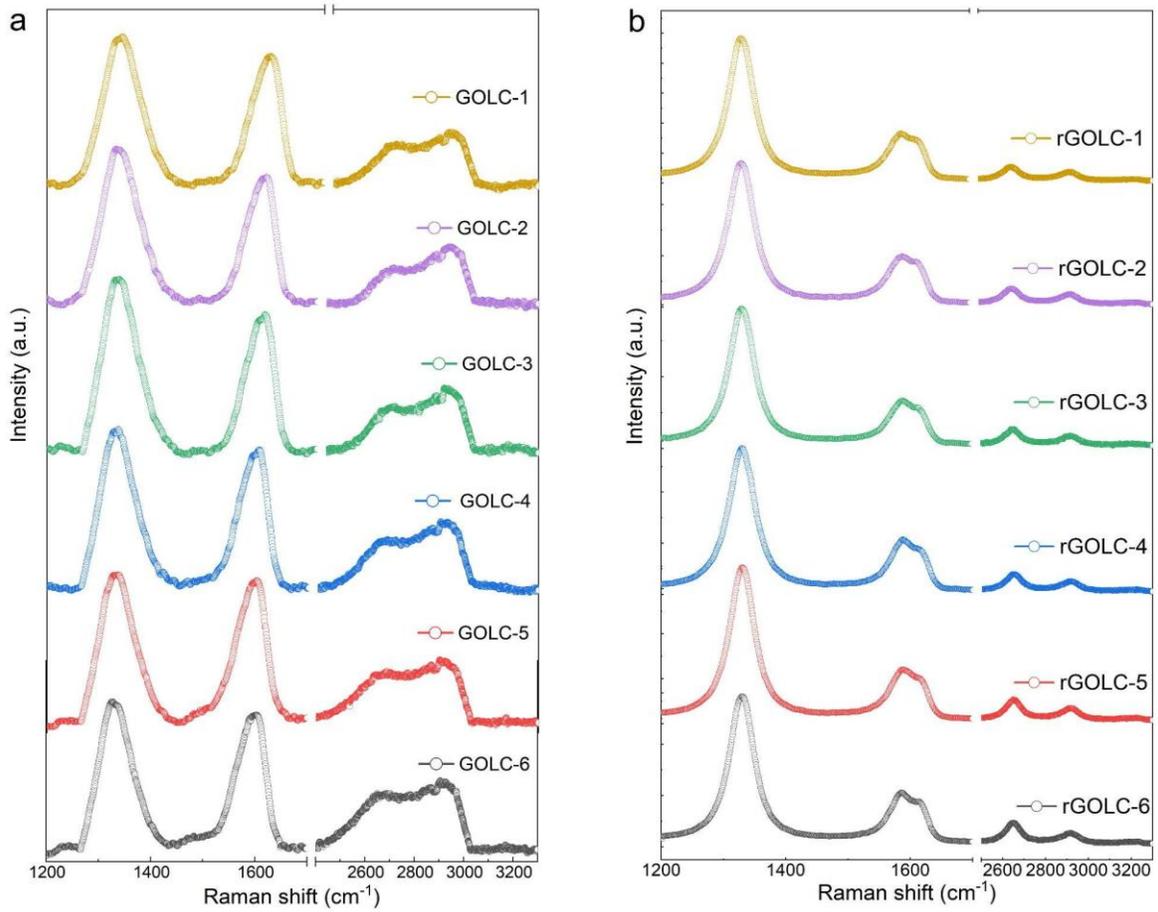

Fig. S5. Raman spectra of GOLC films (a) before and (b) after chemical reduction.

Table S3. Raman band positions for GOLC and rGOLC films.

| Film number | GOLC | rGOLC | GOLC | rGOLC | GOLC | rGOLC | GOLC | rGOLC |
|---|---|---|---|---|---|---|---|---|
| | D, $cm^{-1}$ | | G, $cm^{-1}$ | | 2D, $cm^{-1}$ | | D+G, $cm^{-1}$ | |
| 1 | 1334.7 | 1329.4 | 1594.1 | 1593.8 | 2717.3 | 2649.2 | 2918.1 | 2907.1 |
| 2 | 1335.7 | 1329.3 | 1593.5 | 1593.4 | 2746.6 | 2649.7 | 2923.9 | 2907.9 |
| 3 | 1335.3 | 1329.6 | 1593.1 | 1592.8 | 2726.2 | 2650.3 | 2918.5 | 2908.5 |
| 4 | 1335.6 | 1329.1 | 1592.7 | 1592.5 | 2724.5 | 2650.1 | 2920.5 | 2906.4 |
| 5 | 1335.4 | 1329.7 | 1592.1 | 1593.0 | 2714.7 | 2650.5 | 2917.6 | 2907.4 |
| 6 | 1333.9 | 1329.4 | 1591.8 | 1592.1 | 2712.8 | 2648.4 | 2919.0 | 2904.8 |

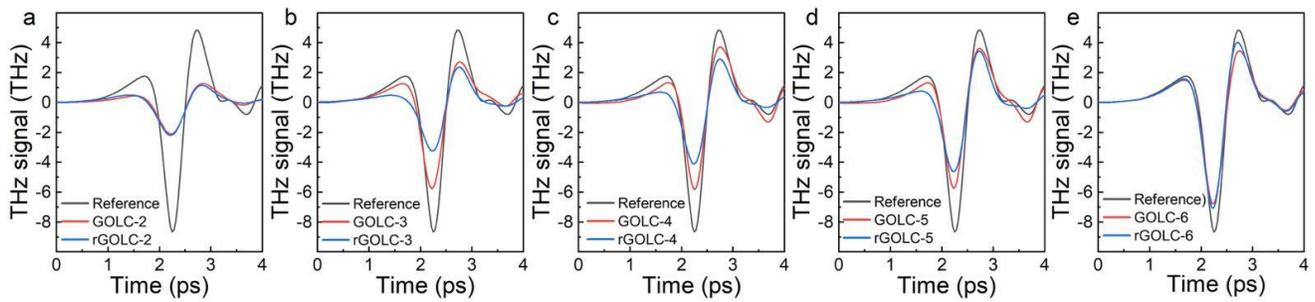

Fig. S6. Temporal profiles of a THz pulse for the GO and rGO LC films numbered from 2 (a) to 6 (e) along with the reference filter.

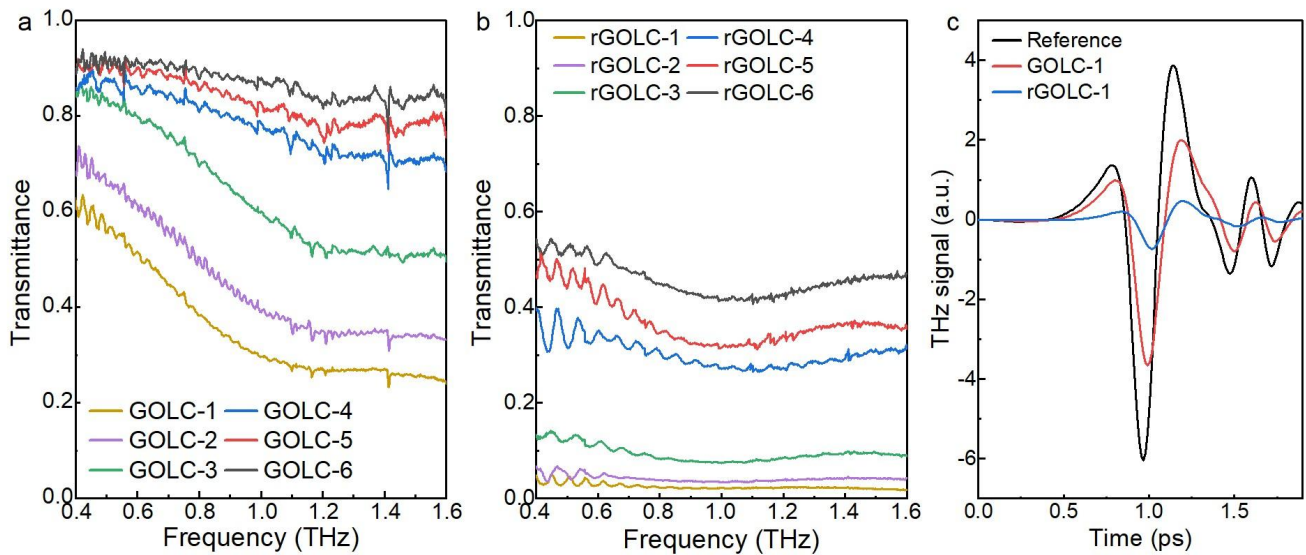

Fig. S7. THz response of GO and rGO LC films. Transmission spectra of the samples with different thicknesses for (a) GOLC and (b) rGOLC films, respectively (for the GOLC films 1 representing the film with 2.12 μm thickness and 6 with 0.28 μm, while for the rGOLC films the corresponding thicknesses are 1.68 μm and 0.16 μm, respectively). (c) Temporal profiles of a THz pulse for the GO and rGO LC films with 2.12 μm and 1.68 μm thicknesses, and the reference filter.